\documentclass[sigconf]{acmart}

\usepackage{booktabs} 
\usepackage{todonotes}
\usepackage{color,soul}
\usepackage{balance}

\copyrightyear{2018} 
\acmYear{2018} 
\setcopyright{licensedusgovmixed}
\acmConference[WPES'18]{2018 Workshop on Privacy in the Electronic Society}{October 15, 2018}{Toronto, ON, Canada}
\acmBooktitle{2018 Workshop on Privacy in the Electronic Society (WPES'18), October 15, 2018, Toronto, ON, Canada}
\acmPrice{15.00}
\acmDOI{10.1145/3267323.3268949}
\acmISBN{978-1-4503-5989-4/18/10}
\acmYear{2018}
\copyrightyear{2018}
\editor{Jennifer B. Sartor}
\editor{Theo D'Hondt}
\editor{Wolfgang De Meuter}

\begin{document}
\title{Issues Encountered Deploying Differential Privacy}

\author{Simson L. Garfinkel}
\orcid{0000-0003-1294-2831}
\affiliation{%
  \institution{US Census Bureau}
  \streetaddress{4600 Silver Hill Road}
  \city{Suitland}
  \state{MD}
  \postcode{20746}
}
\email{simson.l.garfinkel@census.gov}

\author{John M. Abowd}
\orcid{0000-0002-0998-4531}
\affiliation{%
  \institution{US Census Bureau}
  \streetaddress{4600 Silver Hill Road}
  \city{Suitland}
  \state{MD}
  \postcode{20746}
}
\email{john.maron.abowd@census.gov}

\author{Sarah Powazek}
\orcid{}
\affiliation{%
  \institution{MIT}
  \streetaddress{77 Mass. Ave}
  \city{Cambridge}
  \state{MA}
  \postcode{02139}
}
\email{powazek@mit.edu}


\begin{abstract}
When differential privacy was created more than a decade ago,  the
motivating example was statistics published by an official
statistics agency. In attempting to transition differential
privacy from the academy to practice, the U.S. Census Bureau has
encountered many challenges unanticipated by differential privacy's
creators. These challenges include
obtaining qualified personnel and a suitable computing environment,
the difficulty accounting for all uses of the confidential data,
the lack of release mechanisms that align with the needs of data users,
the expectation on the part of data users that they will have access
to micro-data,
and the difficulty in setting the value of the privacy-loss parameter,
$\epsilon$ (epsilon), and 
the lack of  tools and trained individuals to
verify the correctness of differential privacy implementations.
\end{abstract}

%
%


\maketitle
\section{Introduction}
The U.S. Census Bureau is the largest  agency in
the Federal Statistical System. According to the Census Bureau's mission
statement, ``The Census Bureau's mission is to serve
as the leading source of quality data about the nation's people and
economy. We honor privacy, protect confidentiality, share our
expertise globally, and conduct our work openly.''\cite{mission}

As the 2020 Census approaches, focus turns to the Census Bureau as it deploys differential privacy to protect privacy in the upcoming decennial census. Invented by Dwork et.\ al in 2006, differential privacy provides a mathematical definition for the privacy loss to individuals associated with the publishing of statistics based on their confidential data. Today the differential privacy literature provides numerous mechanisms for privacy preserving data publishing and privacy preserving data mining while limiting the resulting privacy loss to mathematically provable bounds\cite{dpbook}. 
 
The 2020 Census data processing system begins by attempting to collect data from all people living in the United States through a variety of means, including an online instrument, a telephone voice-response system, a form that can be mailed in, and ``enumerators'' who travel from house-to-house for non-response follow-up (NRFU)\cite{csac-admin-modeling}. These confidential data will collected and processed to create the Census Unedited File (CUF), which will contain a block-by-block list of every person in the United States. These data must be completed in time to meet the statutory deadline for reapportioning the House of Representatives (December 31, 2020). 
Subject matter experts working with Census-developed software review the CUF and make corrections based on their expertise and other data sources. The result is the Census Edited File (CEF). The Disclosure Avoidance System (DAS), currently under development, will use a novel differential privacy mechanism to add noise to the CEF, producing the Microdata Detail File (MDF) that the Census Bureau's tabulation system will use to create the traditional data products. 

In 2008, the Census Bureau deployed OnTheMap, the first production system to use differential privacy\cite{onthemap}. Six years later, Google deployed RAPPOR\cite{Erlingsson:2014:RRA:2660267.2660348}, the second major production system to use differential privacy, in its Chrome web browser. Today, differential privacy is also being used by Apple\cite{apple-dp} and Microsoft\cite{microsoft-dp}. Although these examples all use differential privacy to protect data supplied by individuals, they use it in different ways, for different purposes. The Census Bureau operates as a \emph{trusted  curator}, which collects sensitive data from individuals, performs statistical tabulations, and publishes them. Trusted curators use differential privacy to prevent matching between a respondent's identity, their data, and a specific data release, which is the Census Bureau's legal requirement under Section 9 of the Census Act, U.S. Code Title 13. Google, Apple and Microsoft use the \emph{local model} of differential privacy: randomization is performed by software running on the individual's computer. These companies use differential privacy so that they cannot make reliable inferences about specific users. These companies use differential privacy to increase public acceptance of their data collection methods.

In 2017, the Census Bureau announced that it would be using differential privacy as the privacy protection mechanism for the 2020 Census of Population of Housing\cite{census-csac-2018}. There is no off-the-shelf mechanism for applying differential privacy to a national census. Although in principle, the Census Bureau could apply Google's
RAPPOR mechanism to the raw census returns, any resulting tabulations would contain far too much noise for any
sensible value of $\epsilon$ to be of much statistical value. To use the Census Bureau's terminology,  the resulting statistics would likely not meet ``fitness for use'' standards, which are also part of the mandate in the Census Act. The same result would ensue if the Census Bureau employed the original Laplace Mechanism\cite{Dwork:2006:CNS:2180286.2180305} to protect its publication tables. An added complication of the Laplace Mechanism is that the tables would not be internally consistent, which might create concerns for data users. Instead, the Census Bureau revealed that it was developing, implementing, testing, and deploying a new differential privacy mechanism. It committed to publishing the mechanism in the peer-reviewed academic literature and making the implementation available to the public, along with suitable test data.

Surprisingly, the Census Bureau's experience with OnTheMap did not significantly prepare the organization for the difficulty of deploying differential privacy for the 2020 Census. OnTheMap was a new product that was designed to incorporate modern privacy protection. In comparison, the decennial Census of Population and Housing, first performed under the direction of Thomas Jefferson in 1790, is the oldest and most expensive statistical undertaking of the U.S. government. Transitioning existing data products to differential privacy has revealed both today's limits in the field of formal privacy, and demonstrated the difficulty of retrofitting legacy statistical products to conform with modern privacy practice.

\section{Prior Work}
Statistical agencies of the U.S. government have traditionally used 
statistical disclosure limitation techniques\cite{wp22} to protect confidentiality; \citeauthor{lauger2014} details how those
techniques were applied to many data products released by the U.S. Census Bureau\cite{lauger2014}.

\citeauthor{abowd2016-why} identifies the challenges faced by statistical agencies in reconciling their traditional disclosure limitation practices with the modern realities of database reconstruction\cite{abowd2016-why,abowd2017-how}, which is made possible because of the large number of statistics published by official statistical agencies,  the availability of large scale computational resources, and third-party data that can  improve the accuracy of the reconstructed database when used in a re-identification attack.

\citeauthor{abowd-new} proposes an approach that statistical agencies can use to set $\epsilon$ using economic theory\cite{abowd-new}.

\section{Specific Challenges}

Here we present some of the  challenges that the Census Bureau has encountered during the deployment of differential privacy. We group the challenges into those that arise from current limitations in the mathematics of differential privacy, those resulting from operational complexities within the Census Bureau, and issues faced by the agency's data users.

\subsection{Scientific Issues}

Differential privacy is less than 15 years old, and most existing mechanisms were created for computer science applications, not the needs of official statistical agencies.  

\emph{Hierarchical Mechanisms.} For the 2020 Census, the agency desired a mechanism that controlled the error as statistics were reported from smaller geographies (e.g. blocks and block groups) to larger geographies (e.g. census tracts and counties) such that the  error would decrease as the population in the relevant geography increased. This required the Census Bureau to develop a set of novel hierarchical mechanisms designed to optimize the accuracy of multiple queries simultaneously.

\emph{Invariants.} For the 2018 End-to-End test, the Census Bureau is reporting exact counts for some statistics (e.g. the number of people living on each block) but privatized counts for other statistics (e.g. the number of Hispanics living on each block). The agency has adopted the term \emph{invariants} to describe statistics that are not changed by the application of differential privacy, and views them as restrictions on the universe of neighboring databases. Nevertheless, there is no well-developed theory for how differential privacy operates in the presence of such invariants. In addition, the historical reasons for having invariants may no longer be consistent with the Census Bureau's confidentiality mandate.

\emph{Stratified sampling.} Between 1940 and 2000, the Census Bureau used two census forms: a \emph{short-form} sent to the majority of households, and a \emph{long-form} with more questions that was sent to a subset. In 2005 the Bureau replaced the long-form with the American Community Survey, a project that continuously measures the U.S. population using a stratified probability sample. Currently, there is no accepted mechanism for applying differential privacy to the results of such sampling. This has delayed the introduction of formal privacy mechanisms for the American Community Survey.

\emph{Quality Metrics.} While the trade-off between between statistical accuracy and privacy loss is at the heart of differential privacy, there are many metrics for assessing the quality of a published dataset. One approach is to calculate the $L_{1}$ error between the true data (i.e. without disclosure limitation) and the privatized data. This is a coarse measure: a disclosure limited product with a high $L_{1}$ compared to the same product without disclosure limitation may still be very accurate for its intended use. Ideally, if intended uses are known, they can be incorporated into the privacy mechanism so that the usefulness is higher for the same privacy-loss budget, allowing the overall privacy-loss budget to be better deployed.

\emph{Presenting and Resolving Equity Issues.} Because the Census Bureau intends to publish many tables drawn from the same confidential database and controlled by an overall privacy-loss budget, there is an opportunity to make some tables more accurate at the cost of making other tables less accurate.  These can be thought of as issues of fairness between different consumers of the Census data, which can be described as an \emph{equity issue}. In principle, these issues are no different  from the  decisions that statistical agencies routinely make about allocating a fixed dollar sampling budget among sub-populations in order to obtain estimates that are fit for use on those sub-populations. Differential privacy lacks a well-developed theory for measuring the relative impact of added noise on the utility of different data products, tuning equity trade-offs, and presenting the impact of such decisions. 

\emph{Establishing a Value of Epsilon.} Before the arrival of differential privacy at the Census Bureau, disclosure avoidance had aspects of the black arts. Knowledge of the actual disclosure avoidance techniques and parameters was restricted to a small group of specialists, and the remainder of the agency treated disclosure avoidance as a black box that input dangerous data and output clean, safe data. The proponents of differential privacy, in contrast, have always maintained that the setting of $\epsilon$ is a policy question, not a technical one. When the Census Bureau announced that it was adopting differential privacy, it also stated that the value of $\epsilon$ would be set by policy makers, not  technologists. But how should policy makers do that? Here, the literature of differential privacy is very sparse.

To date, the Census Bureau's Data Stewardship Executive Policy committee (DSEP) has set the values of $\epsilon$ for one data product. The value was set by having the practitioner prepare a set of graphs that showed the trade-off between privacy loss ($\epsilon$) and accuracy. The group then picked a value of $\epsilon$ that allowed for sufficient accuracy, then tripled it, so that the the researchers would be able to make several additional releases with the same data set without having to return to DSEP to get additional privacy-loss budget. The value of $\epsilon$ that was given out  was far higher than those envisioned by the creators of differential privacy. (In their contemporaneous writings, differential privacy's creators clearly imply that they expected values of $\epsilon$ that were ``much less than one.''\cite{privacy-integrated-queries}). 

\emph{Mechanism Development.} More efficient mechanisms and proofs with tighter bounds are needed to lower amounts of noise for the same level of privacy loss, and to make efficient use of the privacy-loss budget for iterative releases of edited and corrected statistics.

\subsection{Operational Issues}

\emph{Obtaining Qualified Personnel and Tools.}
An early problem faced by the Census Bureau was not technical, but operational: it lacked subject matter experts skilled in the theory and practice of differential privacy. 
In part, this is because only a smattering of universities cover the topic of differential privacy in an instructional setting, and then typically only in a single upper-level computer science course. The Census Bureau, in contrast, typically hires graduates with degrees in mathematics, statistics or economics for its ``mathematical statistician'' career tract.  And while there is a now a textbook that covers the theory of differential privacy\cite{dpbook},  reading a textbook does not provide the necessary expertise to develop correct differential privacy algorithms and implementations. The sparsity of expertise was noted by the Bipartisan Commission on Evidence-based Policymaking, which strongly recommended the adoption of privacy-enhancing data analysis tools while recognizing that there was a dearth of existing tools \cite{CEP2017}.

Likewise, there is a profound lack of toolkits for  performing differential privacy calculations and for verifying the correctness of specific implementations. It is now 12 years since the invention of differential privacy: the situation is analogous to the state of Public Key Cryptography in 1989. 
This has impacted both high-profile projects such as the 2020 Census, as well as the day-to-day work involved in producing more than 100 regular data products and supporting hundreds of researchers in the Federal Statistical Research Data Centers.

\emph{Recasting high-sensitivity queries.}
The 2010 Census publications included statistics about individuals, statistics about households, and statistics reflecting the interaction of the two. The sensitivity of most counting queries is 1---for example, a statistic that reports the number of males and females on a block, or the number of households. Some queries that combine these kinds of statistics also have a sensitivity of 1, such as the number of households headed by a female. But some queries have a much higher sensitivity. For example, a query asking the number of children in households headed by a female has a sensitivity equal to the largest permissible household size. An added complication is that this value needs to be specified in advance, as part of the overall design of the survey, rather than derived by looking at the data, lest information about the presence of a specific large family in the survey data be revealed. 

Currently, the DAS team is working with data users to redesign the publication tables, with the hope of lowering their sensitivity. For example, instead of reporting the number of children that are in a household headed by a person who is Hispanic, the Census could  report the number of Hispanic children. It could also protect the original query, but at more aggregated levels of geography.

\emph{Structural Zeros.} Bringing differential privacy to the 2020 Census required in-depth discussions of the difference between \emph{structural zeros} and \emph{sampling zeros}\cite{bishop-fienberg-holland-1974}.
Structural zeros are those enforced by the Census Bureau's edit rules (``there can be no six-year old mothers with 30-year-old children''), while sampling zeros emerge from the data collection effort (``no women over 65 were found living in this facility''). Injected statistical noise can make sampling zeros positive (2 women over 65 are reported living in the facility), but cannot be allowed to undo the edit rules. 

In practice, the distinction between structural and sampling zeros in an operational context is far less clear. For example, is the number of females in a male prison zero because there are none living there (a sampling zero), or because they are prohibited from living there (a structural zero)? For that matter, how should the Census determine that a facility is single-sex? Previously, whether or not a group quarters was a single-sex might have been determined by looking at the data; this is not permissible in a system that implements differential privacy.

\emph{Obtaining a Suitable Computing Environment.}
The algorithms being developed for the 2020 Census require significant post-processing following the application of noise. In order to characterize their behavior, Census Bureau researchers will perform many runs on the algorithms with historical data, requiring at least three
orders of magnitude greater computing resources than were needed for the 2010 Census. Although the Bureau is migrating from on-premise computing to a cloud-based environment, this migration was delayed because of security concerns, resulting in substantial delays in the development of the 2020 DAS.

\emph{Accounting for All Uses of Confidential Data.}
A key feature of the previous disclosure avoidance mechanism was that it did not change the values of many tabulations at high levels of geography. Thus, many reports from the 2000 and 2010 Censuses could be produced using the confidential data and without applying further disclosure avoidance.

A fundamental requirement of differential privacy is that all calculations involving private data must have noise added before they can be made public. As a result, the Census Bureau has had to identify every use of confidential data in the execution and processing of the 2020 Census. New and unanticipated requirements have emerged during the design of the system after the team thought that the design was locked down.

\emph{Lack of Final Specifications.} Beyond those issues arising from the application of differential privacy, the team building the DAS has also faced by the fluid nature of the decennial census. Many of the Census Bureau data products have been traditionally developed near the end of the decade in consultation with the data users. This collaborative process helps ensure the utility of the census data, but it is at odds with the design and development of a differential privacy system, which requires that all computations be known in advance, or that some amount of privacy-loss budget be reserved for future use. 

\subsection{Issues Faced by Data Users}

\emph{Access to Micro-data.}
Many Census Bureau data users are accustomed to using micro-data, like those originally released for the 1960 Census, that are either raw or that have undergone only limited confidentiality edits as part of their disclosure avoidance. Unfortunately, record-level data are exceedingly difficult to protect in a way that offers real privacy protection while leaving the data useful for unspecified analytical purposes. At present, the Census Bureau advises research users who require such data to consider restricted-access modalities\cite{restricted-use-microdata}.

\emph{Difficulties Arising from Increased Transparency.}
Most users of the 2000 and 2010 Censuses were not aware of the details of the disclosure avoidance mechanism nor its impact on their results. With 2020 Census data, users will be aware that noise has been added, and they will be able to calculate the margin of error that the noise introduces. Some data users are confused about this \emph{margin of error}, a term that they traditionally associate with sample surveys. While coverage error has long been openly discussed and analyzed,\cite{g01} 
discussion of the error caused by disclosure avoidance procedures, historically called ``confidentiality edits,'' has been terse and limited to qualitative statements\cite{sf1.pdf}.

\emph{Misunderstandings about Randomness and Noise Infusion.}
A key mechanism of differential privacy is adding random noise to tabulated data before releasing. 
By design, the noise-injection mechanisms used by the Census Bureau will result in increased accuracy as population sizes increase. Explaining this to data users, community leaders and the general public will be critical to the acceptance of this new disclosure avoidance methodology.  

For example, some statistical programmers want to use repeatable random number generators for regression testing and production, and have the ability to re-run the privacy mechanism if the first set of coins produces results that are deemed unacceptable. Differential privacy is clearly incompatible with this notion.

Although there are many technical papers explaining differential privacy, including the Harvard University Privacy Tools Project\cite{tutorial} and the Duke University tutorials\cite{Machanavajjhala:2017:DPW:3035918.3054779}, their academic language is not accessible to many of the Census Bureau's data users. The  lack of simplified materials to promote a general understanding of differential privacy increases the likelihood of misunderstanding. 

\section{Recommendations}
Despite the numerous challenges differential privacy adoption faced, it has taken root in the Census Bureau. Here, we present recommendations for furthering its integration into the Census Bureau and overcoming some of the hurdles outlined above. 

\emph{Repeated Discussions with Decision Makers.} The deployment of differential privacy within the Census Bureau marks a sea change for the way that official statistics are produced and published. But despite the problems encountered, the Census Bureau has not reconsidered its decision to adopt modern disclosure avoidance mechanisms. We believe that this is a result of the Census Bureau's longstanding commitment to confidentiality protections and the adoption of modern methodological techniques. Repeated discussions with both the Census Bureau's governing boards and with data users are vital in assembling and maintaining institutional support for this transformative effort.

\emph{Controlled Vocabulary.} The Census Bureau has found it helpful to establish a controlled vocabulary of terms for discussions of matters involving differential privacy. In computing and mathematics, it is common for practitioners to adopt many different words to mean the same thing (and, conversely, to use the same words to mean different things in different contexts). 
Internal comprehension as well as the ease of communicating with others has been helped by having a controlled vocabulary, enforced from the highest levels of technical management.

\emph{Integrated Communications.} The Census Bureau has created a communications team staffed with senior members of several directorates for the purpose of working with data users and the public on promoting understanding of the new privacy initiative. This team plays a pillar role in the acceptance of differential privacy, both internally and externally to the Census Bureau. With a public-facing educational tutorial forthcoming, and a suite of informative media in the works, they are making user-level understanding of differential privacy rapidly more available to non-experts. 

Finally, the Census Bureau is expanding its educational efforts on the topic of differential privacy. 

\section{Conclusion}
The Census Bureau is now two years into the process of modernizing its disclosure avoidance systems to incorporate formal privacy protection techniques. Although this process has proven to be challenging across disciplines, it promises to reward the efforts of the Census Bureau. In order to attempt privacy protection on the same scale without differential privacy, the Census Bureau could publish dramatically fewer tables and simply hope that they haven't leaked enough information to allow an attacker to perform database reconstruction. By implementing differential privacy, the Census Bureau can mathematically limit the privacy loss associated with each publication. Beyond the 2020 Census, the Census Bureau intends to use differential privacy or related formal privacy systems to protect all of its statistical publications.

It is noteworthy that this institution is not only implementing differential privacy in its statistical analyses, but truly integrating it into its organizational structure. With staff in communications, research, statistics, and computer science familiar with and supportive of differential privacy, a set of diverse employees equipped with privacy tools will be available in the Census Bureau beyond the 2020 Census. The methods put in place for the 2018 and 2020 implementations will act as templates, greatly easing its adoption in future statistical projects. With skilled staff and effective methodology in place, differential privacy can make lasting improvements to privacy protection at the federal government's largest statistical agency. 

\textbf{\small DISCLAIMER: This paper is presented with the hope that its content may be of interest to the general statistical community. The views in these papers are those of the authors, and do not necessarily represent those of the Census Bureau.}
\clearpage
\bibliographystyle{ACM-Reference-Format}
\balance
\bibliography{main}


\begin{thebibliography}{22}


\ifx \showCODEN    \undefined \def \showCODEN     #1{\unskip}     \fi
\ifx \showDOI      \undefined \def \showDOI       #1{#1}\fi
\ifx \showISBNx    \undefined \def \showISBNx     #1{\unskip}     \fi
\ifx \showISBNxiii \undefined \def \showISBNxiii  #1{\unskip}     \fi
\ifx \showISSN     \undefined \def \showISSN      #1{\unskip}     \fi
\ifx \showLCCN     \undefined \def \showLCCN      #1{\unskip}     \fi
\ifx \shownote     \undefined \def \shownote      #1{#1}          \fi
\ifx \showarticletitle \undefined \def \showarticletitle #1{#1}   \fi
\ifx \showURL      \undefined \def \showURL       {\relax}        \fi
\providecommand\bibfield[2]{#2}
\providecommand\bibinfo[2]{#2}
\providecommand\natexlab[1]{#1}
\providecommand\showeprint[2][]{arXiv:#2}

\bibitem[\protect\citeauthoryear{??}{res}{2018}]%
        {restricted-use-microdata}
 \bibinfo{year}{2018}\natexlab{}.
\newblock \bibinfo{title}{Restricted-Use Microdata}.
\newblock
\newblock
\urldef\tempurl%
\url{https://www.census.gov/research/data/restricted_use_microdata.html#CRE1}
\showURL{%
\tempurl}
\newblock
\shownote{Last Accessed July 14, 2018.}


\bibitem[\protect\citeauthoryear{Abowd}{Abowd}{2016}]%
        {abowd2016-why}
\bibfield{author}{\bibinfo{person}{John Abowd}.}
  \bibinfo{year}{2016}\natexlab{}.
\newblock \showarticletitle{Why Statistical Agencies Need to Take Privacy-loss
  Budgets Seriously, and What It Means When They Do}.
\newblock \bibinfo{journal}{\emph{Labor Dynamics Institute}}
  (\bibinfo{date}{Dec. 7} \bibinfo{year}{2016}).
\newblock
\urldef\tempurl%
\url{http://digitalcommons.ilr.cornell.edu/ldi/32/}
\showURL{%
\tempurl}


\bibitem[\protect\citeauthoryear{Abowd}{Abowd}{2017}]%
        {abowd2017-how}
\bibfield{author}{\bibinfo{person}{John Abowd}.}
  \bibinfo{year}{2017}\natexlab{}.
\newblock \showarticletitle{How Will Statistical Agencies Operate When All Data
  Are Private?}
\newblock \bibinfo{journal}{\emph{Journal of Privacy and Confidentiality}}
  \bibinfo{volume}{7} (\bibinfo{year}{2017}).
\newblock
Issue 3.
\urldef\tempurl%
\url{https://doi.org/10.29012/jpc.v7i3.404.}
\showURL{%
\tempurl}


\bibitem[\protect\citeauthoryear{Abowd and Schmutte}{Abowd and Schmutte}{[n.
  d.]}]%
        {abowd-new}
\bibfield{author}{\bibinfo{person}{John~M. Abowd} {and} \bibinfo{person}{Ian~M.
  Schmutte}.} \bibinfo{year}{[n. d.]}\natexlab{}.
\newblock \showarticletitle{An Economic Analysis of Privacy Protection and
  Statistical Accuracy as Social Choices}.
\newblock \bibinfo{journal}{\emph{American Economic Review}}
  (\bibinfo{year}{[n. d.]}).
\newblock
\urldef\tempurl%
\url{https://arxiv.org/abs/1808.06303}
\showURL{%
\tempurl}
\newblock
\shownote{forthcoming.}


\bibitem[\protect\citeauthoryear{Abraham, Haskins, Glied, Groves, Hahn, Hoynes,
  Liebman, Meyer, Haskins, Ohm, Potok, Mosier, Shea, Sweeney, Troske, and
  Wallin}{Abraham et~al\mbox{.}}{2017}]%
        {CEP2017}
\bibfield{author}{\bibinfo{person}{Katherine~G. Abraham}, \bibinfo{person}{Ron
  Haskins}, \bibinfo{person}{Sherry Glied}, \bibinfo{person}{Robert~M. Groves},
  \bibinfo{person}{Robert Hahn}, \bibinfo{person}{Hilary Hoynes},
  \bibinfo{person}{Jeffrey~B. Liebman}, \bibinfo{person}{Bruce~D. Meyer},
  \bibinfo{person}{Ron Haskins}, \bibinfo{person}{Paul Ohm},
  \bibinfo{person}{Nancy Potok}, \bibinfo{person}{Kathleen~Rice Mosier},
  \bibinfo{person}{Robert~J. Shea}, \bibinfo{person}{Latanya Sweeney},
  \bibinfo{person}{Kenneth~R. Troske}, {and} \bibinfo{person}{Kim~R. Wallin}.}
  \bibinfo{year}{2017}\natexlab{}.
\newblock \bibinfo{booktitle}{\emph{The Promise of Evidence-Based
  Policymaking}}.
\newblock \bibinfo{publisher}{Comission on Evidence-Based Policymaking},
  \bibinfo{address}{Washington, DC}.
\newblock
\urldef\tempurl%
\url{https://www.cep.gov/cep-final-report.html}
\showURL{%
\tempurl}


\bibitem[\protect\citeauthoryear{Andersson, Abowd, Graham, Wu, and
  Vilhuber}{Andersson et~al\mbox{.}}{2009}]%
        {onthemap}
\bibfield{author}{\bibinfo{person}{Fredrik Andersson}, \bibinfo{person}{John~M.
  Abowd}, \bibinfo{person}{Matthew Graham}, \bibinfo{person}{Jeremy Wu}, {and}
  \bibinfo{person}{Lars Vilhuber}.} \bibinfo{year}{2009}\natexlab{}.
\newblock \showarticletitle{Formal Privacy Guarantees and Analytical Validity
  of OnTheMap Public-use Data}. In \bibinfo{booktitle}{\emph{Joint
  NSF-Census-IRS Workshop on Synthetic Data and Confidentiality Protection}}.
  \bibinfo{publisher}{Cornell University}, \bibinfo{address}{Suitland, MD}.
\newblock
\urldef\tempurl%
\url{https://ecommons.cornell.edu/handle/1813/47672}
\showURL{%
\tempurl}


\bibitem[\protect\citeauthoryear{Apple Computer}{Apple Computer}{2017}]%
        {apple-dp}
 \bibinfo{year}{2017}\natexlab{}.
\newblock \bibinfo{booktitle}{\emph{Differential Privacy}}.
\newblock \bibinfo{publisher}{Apple Computer}.
\newblock
\urldef\tempurl%
\url{https://www.apple.com/privacy/docs/Differential_Privacy_Overview.pdf}
\showURL{%
\tempurl}


\bibitem[\protect\citeauthoryear{Bishop, Fienberg, and Holland}{Bishop
  et~al\mbox{.}}{1974}]%
        {bishop-fienberg-holland-1974}
\bibfield{author}{\bibinfo{person}{Yvonne~M. Bishop},
  \bibinfo{person}{Stephen~E. Fienberg}, {and} \bibinfo{person}{Paul~W.
  Holland}.} \bibinfo{year}{1974}\natexlab{}.
\newblock \bibinfo{booktitle}{\emph{Discrete Multivariate Analysis: Theory and
  Practice}}.
\newblock \bibinfo{publisher}{Springer}.
\newblock
\urldef\tempurl%
\url{https://www.springer.com/us/book/9780387728056}
\showURL{%
\tempurl}


\bibitem[\protect\citeauthoryear{Ding, Kulkarni, and Yekhanin}{Ding
  et~al\mbox{.}}{2017}]%
        {microsoft-dp}
\bibfield{author}{\bibinfo{person}{Bolin Ding}, \bibinfo{person}{Jana
  Kulkarni}, {and} \bibinfo{person}{Sergey Yekhanin}.}
  \bibinfo{year}{2017}\natexlab{}.
\newblock \bibinfo{title}{Collecting telemetry data privately}.
\newblock
\newblock
\urldef\tempurl%
\url{https://www.microsoft.com/en-us/research/blog/collecting-telemetry-data-privately/}
\showURL{%
\tempurl}


\bibitem[\protect\citeauthoryear{Dwork, McSherry, Nissim, and Smith}{Dwork
  et~al\mbox{.}}{2006}]%
        {Dwork:2006:CNS:2180286.2180305}
\bibfield{author}{\bibinfo{person}{Cynthia Dwork}, \bibinfo{person}{Frank
  McSherry}, \bibinfo{person}{Kobbi Nissim}, {and} \bibinfo{person}{Adam
  Smith}.} \bibinfo{year}{2006}\natexlab{}.
\newblock \showarticletitle{Calibrating Noise to Sensitivity in Private Data
  Analysis}. In \bibinfo{booktitle}{\emph{Proceedings of the Third Conference
  on Theory of Cryptography}} \emph{(\bibinfo{series}{TCC'06})}.
  \bibinfo{publisher}{Springer-Verlag}, \bibinfo{address}{Berlin, Heidelberg},
  \bibinfo{pages}{265--284}.
\newblock
\showISBNx{3-540-32731-2, 978-3-540-32731-8}
\urldef\tempurl%
\url{https://doi.org/10.1007/11681878_14}
\showDOI{\tempurl}


\bibitem[\protect\citeauthoryear{Dwork and Roth}{Dwork and Roth}{2014}]%
        {dpbook}
\bibfield{author}{\bibinfo{person}{Cynthia Dwork} {and} \bibinfo{person}{Aaron
  Roth}.} \bibinfo{year}{2014}\natexlab{}.
\newblock \showarticletitle{The Algorithmic Foundations of Differential
  Privacy}. In \bibinfo{booktitle}{\emph{Foundations and Trends in Theoretical
  Computer Science}}, Vol.~\bibinfo{volume}{9}. \bibinfo{publisher}{NOW},
  \bibinfo{pages}{211--407}.
\newblock


\bibitem[\protect\citeauthoryear{Erlingsson, Pihur, and Korolova}{Erlingsson
  et~al\mbox{.}}{2014}]%
        {Erlingsson:2014:RRA:2660267.2660348}
\bibfield{author}{\bibinfo{person}{\'{U}lfar Erlingsson},
  \bibinfo{person}{Vasyl Pihur}, {and} \bibinfo{person}{Aleksandra Korolova}.}
  \bibinfo{year}{2014}\natexlab{}.
\newblock \showarticletitle{RAPPOR: Randomized Aggregatable Privacy-Preserving
  Ordinal Response}. In \bibinfo{booktitle}{\emph{Proceedings of the 2014 ACM
  SIGSAC Conference on Computer and Communications Security}}
  \emph{(\bibinfo{series}{CCS '14})}. \bibinfo{publisher}{ACM},
  \bibinfo{address}{New York, NY, USA}, \bibinfo{pages}{1054--1067}.
\newblock
\showISBNx{978-1-4503-2957-6}
\urldef\tempurl%
\url{https://doi.org/10.1145/2660267.2660348}
\showDOI{\tempurl}


\bibitem[\protect\citeauthoryear{FCSM}{FCSM}{2005}]%
        {wp22}
FCSM \bibinfo{year}{2005}\natexlab{}.
\newblock \bibinfo{booktitle}{\emph{Working Paper 22: Report on Statistical
  Disclosure Limitation Methodology}}.
\newblock \bibinfo{type}{{T}echnical {R}eport}. \bibinfo{institution}{Federal
  Committee on Statistical Methodology}.
\newblock
\urldef\tempurl%
\url{https://fcsm.sites.usa.gov/reports/policy-wp/}
\showURL{%
\tempurl}


\bibitem[\protect\citeauthoryear{Garfinkel}{Garfinkel}{2018}]%
        {census-csac-2018}
\bibfield{author}{\bibinfo{person}{Simson~L. Garfinkel}.}
  \bibinfo{year}{2018}\natexlab{}.
\newblock \bibinfo{title}{Modernizing Disclosure Avoidance: Report on the 2020
  Disclosure Avoidance System as Implemented for the 2018 End-to-End Test}.
\newblock
\newblock
\urldef\tempurl%
\url{https://www.census.gov/about/cac/sac/meetings/2017-09-meeting.html}
\showURL{%
\tempurl}


\bibitem[\protect\citeauthoryear{Lauger, Wisniewski, and McKenna}{Lauger
  et~al\mbox{.}}{2014}]%
        {lauger2014}
\bibfield{author}{\bibinfo{person}{Amy Lauger}, \bibinfo{person}{Billy
  Wisniewski}, {and} \bibinfo{person}{Laura McKenna}.}
  \bibinfo{year}{2014}\natexlab{}.
\newblock \bibinfo{booktitle}{\emph{Disclosure Avoidance Techniques at the U.S.
  Census Bureau: Current Practices and Research}}.
\newblock \bibinfo{type}{{T}echnical {R}eport}. \bibinfo{institution}{U.S.
  Census Bureau}.
\newblock
\urldef\tempurl%
\url{{https://www.census.gov/srd/CDAR/cdar2014-02\_Discl\_Avoid\_Techniques.pdf}}
\showURL{%
\tempurl}


\bibitem[\protect\citeauthoryear{Machanavajjhala, He, and Hay}{Machanavajjhala
  et~al\mbox{.}}{2017}]%
        {Machanavajjhala:2017:DPW:3035918.3054779}
\bibfield{author}{\bibinfo{person}{Ashwin Machanavajjhala}, \bibinfo{person}{Xi
  He}, {and} \bibinfo{person}{Michael Hay}.} \bibinfo{year}{2017}\natexlab{}.
\newblock \showarticletitle{Differential Privacy in the Wild: A Tutorial on
  Current Practices \&\#38; Open Challenges}. In
  \bibinfo{booktitle}{\emph{Proceedings of the 2017 ACM International
  Conference on Management of Data}} \emph{(\bibinfo{series}{SIGMOD '17})}.
  \bibinfo{publisher}{ACM}, \bibinfo{address}{New York, NY, USA},
  \bibinfo{pages}{1727--1730}.
\newblock
\showISBNx{978-1-4503-4197-4}
\urldef\tempurl%
\url{https://doi.org/10.1145/3035918.3054779}
\showDOI{\tempurl}


\bibitem[\protect\citeauthoryear{McSherry}{McSherry}{2009}]%
        {privacy-integrated-queries}
\bibfield{author}{\bibinfo{person}{Frank McSherry}.}
  \bibinfo{year}{2009}\natexlab{}.
\newblock \showarticletitle{Privacy Integrated Queries}, In
  \bibinfo{booktitle}{Proceedings of the 2009 ACM SIGMOD International
  Conference on Management of Data (SIGMOD)}.
\newblock
\urldef\tempurl%
\url{https://www.microsoft.com/en-us/research/publication/privacy-integrated-queries/}
\showURL{%
\tempurl}


\bibitem[\protect\citeauthoryear{Mule}{Mule}{2012}]%
        {g01}
\bibfield{author}{\bibinfo{person}{Thomas Mule}.}
  \bibinfo{year}{2012}\natexlab{}.
\newblock \bibinfo{booktitle}{\emph{Census Coverage Measurement Estimation
  Report: Summary of Estimates of Coverage for Persons in the United States}}.
\newblock \bibinfo{type}{{T}echnical {R}eport}. \bibinfo{institution}{U.S.
  Census Bureau}.
\newblock
\urldef\tempurl%
\url{https://www.census.gov/coverage_measurement/pdfs/g01.pdf}
\showURL{%
\tempurl}


\bibitem[\protect\citeauthoryear{Nissim, Steinke, Wood, Altman, Bembenek, Bun,
  Gaboardi, O'Brien, and Vadhan}{Nissim et~al\mbox{.}}{2018}]%
        {tutorial}
\bibfield{author}{\bibinfo{person}{Kobbi Nissim}, \bibinfo{person}{Thomas
  Steinke}, \bibinfo{person}{Alexandra Wood}, \bibinfo{person}{Micah Altman},
  \bibinfo{person}{Aaron Bembenek}, \bibinfo{person}{Mark Bun},
  \bibinfo{person}{Marco Gaboardi}, \bibinfo{person}{David O'Brien}, {and}
  \bibinfo{person}{Salil Vadhan}.} \bibinfo{year}{2018}\natexlab{}.
\newblock \showarticletitle{Differential Privacy: A Primer for a Non-technical
  Audience (Preliminary Version).}
\newblock \bibinfo{journal}{\emph{Vanderbilt Journal of Entertainment and
  Technology Law}} (\bibinfo{year}{2018}).
\newblock
\newblock
\shownote{Forthcoming.}


\bibitem[\protect\citeauthoryear{U.S. Census Bureau}{U.S. Census
  Bureau}{2012}]%
        {sf1.pdf}
U.S. Census Bureau \bibinfo{year}{2012}\natexlab{}.
\newblock \bibinfo{booktitle}{\emph{2010 Census Summary File 1: 2010 Census of
  Population and Housing, Technical Documentation}}.
\newblock \bibinfo{type}{{T}echnical {R}eport}. \bibinfo{institution}{U.S.
  Census Bureau}.
\newblock
\urldef\tempurl%
\url{https://www.census.gov/prod/cen2010/doc/sf1.pdf}
\showURL{%
\tempurl}


\bibitem[\protect\citeauthoryear{US Census Bureau}{US Census Bureau}{2017}]%
        {csac-admin-modeling}
US Census Bureau \bibinfo{year}{2017}\natexlab{}.
\newblock \bibinfo{booktitle}{\emph{Administrative Records Modeling Update for
  the Census Scientific Advisory Committee}}.
\newblock \bibinfo{type}{{T}echnical {R}eport}. \bibinfo{institution}{US Census
  Bureau}.
\newblock
\urldef\tempurl%
\url{https://www2.census.gov/cac/sac/meetings/2017-03/admin-records-modeling.pdf}
\showURL{%
\tempurl}


\bibitem[\protect\citeauthoryear{{U.S. Census Bureau}}{{U.S. Census
  Bureau}}{2017}]%
        {mission}
\bibfield{author}{\bibinfo{person}{{U.S. Census Bureau}}.}
  \bibinfo{year}{2017}\natexlab{}.
\newblock \bibinfo{title}{Our Mission}.
\newblock
\newblock
\urldef\tempurl%
\url{https://www.census.gov/about/what.html}
\showURL{%
\tempurl}


\end{thebibliography}

\end{document}